\RequirePackage{fix-cm}
\documentclass[twocolumn]{svjour3}          
\smartqed  
\usepackage{graphicx}
\journalname{Journal of Supercomputing}
\usepackage{amssymb}
\usepackage[normalem]{ulem}
\usepackage[fleqn]{amsmath}
\usepackage[varg]{txfonts}
\usepackage{color}
\usepackage{multirow}
\usepackage{lscape}
\usepackage{soul}
\usepackage[ruled,linesnumbered]{algorithm2e}
\definecolor{pink}{rgb}{1,0.5,0.5}
\usepackage{amsmath}
\usepackage{amsfonts}
\usepackage{amssymb}
\usepackage{varwidth}
\usepackage{enumitem}
\usepackage{tikz}
\usetikzlibrary{calc}
\usetikzlibrary{positioning}
\usetikzlibrary{arrows}
\usetikzlibrary{decorations.pathreplacing}
\usetikzlibrary{fit}
\usetikzlibrary{matrix}
\tikzset{
    vertex/.style = {
        circle,
        fill            = black,
        outer sep = 2pt,
        inner sep = 1pt,
    },
    boxnode/.style = {align=center,draw}
}
\tikzstyle{arrow}=[draw, -latex,solid,line width=0.5pt]
\graphicspath{{./figures/}}
\begin{document}

\title{A Low-overhead Soft-Hard Fault Tolerant Architecture, Design and Management Scheme for Reliable High-performance Many-core 3D-NoC Systems}

\titlerunning{A Low-overhead Soft-Hard Fault-Tolerant Architecture and Management Scheme for Reliable...}        

\author{Khanh N. Dang        \and Michael Meyer \and Yuichi Okuyama \and
        Abderazek Ben Abdallah 
}


\institute{Khanh N. Dang    \and Micheal Meyer \and Yuichi Okuyama \and
    Abderazek Ben Abdallah \at
Adaptive Systems Laboratory\\
Graduate School of Computer Science and Engineering\\
The University of Aizu\\
Aizu-Wakamatsu City, Fukushima 965-8580, Japan\\
              \email{d8162103, benab@u-aizu.ac.jp}           
\and
}

\date{The final publication is available at Springer via https://doi.org/10.1007/s11227-016-1951-0}

\maketitle

\begin{abstract}
The Network-on-Chip (NoC) paradigm has been proposed as a favorable solution to handle the strict communication requirements between the increasingly large number of cores on a single chip. However, NoC systems are exposed to the aggressive scaling down of transistors, low operating voltages, and high integration and power densities, making them vulnerable to permanent (hard) faults and transient (soft) errors. 
A hard fault in a NoC can lead to external blocking, causing congestion across the whole network. A soft error is more challenging because of its silent data corruption, which leads to a large area of erroneous data due to error propagation, packet re-transmission, and deadlock.  
In this paper, we present the architecture and design of a comprehensive soft error and hard fault tolerant 3D-NoC system, named 3D-Hard-Fault-Soft-Error-Tolerant-OASIS-NoC (3D-FETO)\footnote{This project is partially supported by Competitive Research Funding (CRF), The University of Aizu, Reference P-11 (2016), and 
JSPS KAKENHI Grant Number JP30453020}. 
With the aid of efficient mechanisms and algorithms, 3D-FETO is capable of detecting and recovering from soft errors which occur in the routing pipeline stages and leverages reconfigurable components to handle permanent faults in links, input buffers, and crossbars. In-depth evaluation results show that the 3D-FETO system is able to work around different kinds of hard faults and soft errors, ensuring graceful performance degradation, while minimizing additional hardware complexity and remaining power-efficient. 
\end{abstract}

\section{Introduction}
Global interconnects are becoming the principal performance bottleneck for high performance Systems-on-Chip (SoCs)~\cite{Ben2013BOOK}. The 3-dimensional Networks-on-Chip (3D-NoCs) have been proposed as a promising architecture that combines the high parallelism of Network-on-Chip paradigm with the high performance and lower interconnect power of 3-dimensional integration circuits (3D-ICs)~\cite{BenAhmed2013Architecture}. In the past few years, the benefits of 3D Integrated Circuits (3D-ICs) and mesh-based Network-on-Chips (NoCs) have been fused into a promising architecture opening a new horizon for IC design. The parallelism of NoCs can be enhanced in the third dimension thanks to the short wire length and low power consumption of the interconnects of 3D-ICs. As a result, the 3D-NoC paradigm is considered to be one of the most advanced and auspicious architectures for the future of IC design, as it is capable of providing extremely high bandwidth and low power interconnects.  

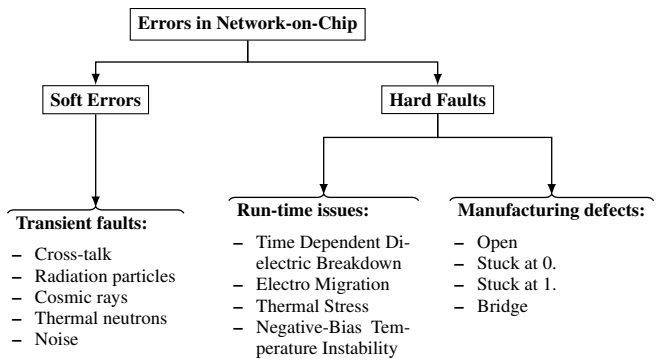
\begin{figure}
    \centering
    \begin{tikzpicture}[font=\scriptsize]
    \node[boxnode] (NoC) at (4,1) {\textbf{Errors in Network-on-Chip}};
    \node[boxnode] (SoftError) at (2,0) {\textbf {Soft Errors}};
    \draw[arrow] (NoC)--++(0,-0.5)-|(SoftError);

    \node (transient) at (2,-2.4) {
        {\begin{varwidth}{\linewidth}
            \textbf{Transient faults:}
            \begin{itemize}[leftmargin=0.1in]
            \item Cross-talk
            \item Radiation particles
            \item Cosmic rays
            \item Thermal neutrons
            \item Noise
            \end{itemize}\end{varwidth}}
    };
    
    \draw[decoration={brace,raise=5pt},decorate,line width=0.5pt]
    ([yshift=-7pt]transient.north west) -- ([yshift=-7pt]transient.north east);
    \draw[arrow] (SoftError.south)--(transient.north);
    
    \node[boxnode] (HardError) at (6.5,0) {\textbf{Hard Faults}};
    \draw[arrow]   (NoC)--++(0,-0.5)-|(HardError);
    \node (run-time) at (5,-2.4) {
        {\begin{varwidth}{10em}
            \textbf{Run-time issues:}
            \begin{itemize} [leftmargin=0.1in]
            \item Time Dependent Dielectric Breakdown
            \item Electro Migration
            \item Thermal Stress
            \item Negative-Bias Temperature Instability
            \end{itemize}\end{varwidth}}
    };
    \draw[decoration={brace,raise=5pt},decorate,line width=0.5pt]
    ([yshift=-7pt]run-time.north west) -- ([yshift=-7pt]run-time.north east);
    \node (manu) at (8,-2.12) {
        {\begin{varwidth}{\linewidth}
            \textbf{Manufacturing defects:}
            \begin{itemize}[leftmargin=0.1in]
            \item Open
            \item Stuck at 0.
            \item Stuck at 1.
            \item Bridge
            \end{itemize}\end{varwidth}}
    };
    \draw[decoration={brace,raise=5pt},decorate,line width=0.5pt]
    ([yshift=-7pt]manu.north west) -- ([yshift=-7pt]manu.north east);
    \draw[arrow]   (HardError)--++(0,-0.5)-|(run-time);
    \draw[arrow]   (HardError)--++(0,-0.5)-|(manu);
    \end{tikzpicture}
    \caption{Taxonomy of errors and faults in NoCs.}
    \label{fig:faults-graph}
\end{figure}
While the NoC paradigm has been increasing in popularity with several commercial chips~\cite{BenAbdallah2006Basic}, 
it is threatened by the decreasing reliability of aggressively scaled transistors. Transistors are approaching the fundamental limits of scaling. Gate widths are nearing the molecular scale, resulting in breakdown and wear out in end products~\cite{Fick2009highly,Karl2006Reliability}. Moreover, the anticipated fabrication geometry in 2018 scales down to $8nm$ with a projected 0.6V supply voltage~\cite{ITRS2012}. In the $8nm$ process, a higher rate of soft errors affect control logic and buffers of NoC routers, leading to chip failure. In addition, the low supply voltage enforces a very narrow noise margin, which makes the architecture vulnerable and sensitive to faults. {As reported in}~\cite{Dixit2011impact}, {the soft error rate increases about 30\% for each 100 \textit{mV} decrease in the supply voltage.} With rising power density and non-ideal threshold and supply voltage scaling, soft errors have become increasingly common during a chip's lifetime~\cite{Eghbal2015Analytical}.  Figure~\ref{fig:faults-graph} shows a detailed taxonomy of different types of error and fault sources in NoCs. {We categorized the faults into two classes: \textit{Hard Faults} and \textit{Soft Errors}.}

Hard faults, including both permanent faults and intermittent faults, can occur during the manufacturing stage or under specific operating circumstances. Intermittent faults periodically occur during operation and can disappear after a certain time. Because these faults do not permanently damage a given component, it can pass through several testing stages, but can still cause  operation failures. Although intermittent faults can disappear after a specific period of time, their inconsistency can be treated as permanent faults to avoid complex situations. For both permanent and intermittent faults, the most natural solution is using redundant components~\cite{DeOrio2012reliable,Constantinides2006Bulletproof}.

Soft errors arise from energetic particles, such as alpha particles and neutrons from cosmic rays, generating electron-hole pairs as they pass through a device. A sufficient amount of accumulated charge may invert the state of a logic device such as a: latch, gate, or SRAM cell; thereby introducing a logic fault into the NoC's operation.
Soft errors do not permanently defect the gate and only occur over a short period of time. Because of their special characteristics, they are unpredictable and unavoidable. Unlike permanent and intermittent faults, transient faults cannot be fixed by replacing the affected components. Instead, they can be recovered by repeating the erroneous operation. A transient failure inside the data path can also be fixed by using code-based techniques (e.g., Error Correction Code (ECC)~\cite{Bertozzi2005Error}). 
Statistically, transient faults are the most common kind of fault accounting for 80\% of failures, as reported in~\cite{Lehtonen2007Online}. Therefore, without an efficient protection mechanism, these errors can compromise the system's functionality and reliability.

Hard fault handling schemes are based on two main approaches: (a) fault-tolerant routing algorithms, which enable packets to avoid faulty nodes in the network~\cite{BenAhmed2013Architecture,DeOrio2012reliable}; (b) architecture-based methods, which use hardware (components) redundancy and/or reconfiguration to recover from faults~\cite{DeOrio2012reliable,Constantinides2006Bulletproof,BenAhmed2016Adaptive}. 
Soft error recovery is  also solved by two main schemes: (a) data corruption handling using Error Correction Code (ECC) based methods~\cite{Lin1984Automatic,Bertozzi2005Error,Yu2010Transient} ; (b) control logic handling using temporal redundancy based methods~\cite{Ernst2003Razor,Yu2013Addressing,Dang2015Soft}. 

Although many researchers have proposed solutions for various individual aspects of on-chip reliability, a comprehensive approach encompassing both soft errors and hard faults pertaining to NoC reliability has yet to evolve. 
In addition, the error  detection  and  diagnosis in NoC  architectures  has  been studied thoroughly in the scope of offline testing; however, with soft errors and intermittent faults becoming a dominant failure mode in modern NoCs and general VLSI systems, a widespread deployment of online test approaches has become crucial.  
In this paper, we present a comprehensive soft error and hard fault tolerant 3D-NoC architecture, named 3D-Hard-Fault-Soft-Error-Tolerant-OASIS-NoC (3D-FETO). With the aid of efficient mechanisms and algorithms, 3D-FETO is capable of detecting and recovering from soft errors occurring in the routing pipeline stages and leverages reconfigurable components to handle permanent fault occurrences in links, input-buffers, and crossbars. The main contributions of this work are summarized as follows: 

\begin{itemize}
    \item A new adaptive 3D router architecture based on a robust hardware  reconfiguration mechanism of the most susceptible components to hardware faults, and on a low-cost method that is capable of detecting and recovering from soft errors in the router pipeline stages. 
    \item An  efficient  scheme  for  online  control  fault  detection  and  diagnosis  in  3D-NoC  systems. 
\end{itemize}

The organization of this paper is as follows: in Section~\ref{sec:related-works}, we present related works.
Section~\ref{sec:arch} presents the adaptive router architecture (SHER-3DR).
In Section~\ref{ssec:DDRM}, we present comprehensive techniques which include fault detection, diagnosis and recovery. 
%
Section~\ref{sec:eva} provides the implementation and evaluation results. Finally, we present the conclusion and our ideas for future work in the last section. 

\section{Related Works} \label{sec:related-works}
A lot of works have addressed the fault-tolerance and reliability issues in NoC architectures. 
In \cite{BenAhmed2013Architecture,BenAhmed2016Adaptive,BenAhmed2014Graceful}, we covered some well-known solutions presented to tackle hard faults; therefore, in this section we mainly focus on solutions related to soft error recovery. As depicted in Table \ref{tab:FT-graph}, they are classified into methods focusing on the Data Path (DP) and methods focusing on the Control Logic (CL) of the router.

\begin{table*}[htbp]
    \begin{center}
        \caption{{Taxonomy of different error recovery protocols and architectures in NoCs.}}
        \label{tab:FT-graph}
        \begin{tabular}{|c|c|l|}
            \hline
            \textbf{Fault Type} & \textbf{Position/Method} &  \textbf{Fault Tolerant Method} \\ \hline \hline
            \multirow{5}{*}{Soft Errors} & \multirow{2}{*}{ Data Path} & Automatic Re-transmission Request \cite{Lin1984Automatic} \\ \cline{3-3}
                                          &           & Error Detecting/Correcting Code \cite{Bertozzi2005Error,Yu2010Transient} \\ \cline{2-3} \cline{2-3}
                                          & \multirow{3}{*}{Control Logic} & Logic/Latch Hardening \cite{Ernst2003Razor,Radetzki2013Methods} \\ \cline{3-3}
                                          &           &Pipeline Redundancy \cite{Dang2015Soft}\\ \cline{3-3}
                                          &           &Monitoring and Correcting model \cite{Yu2013Addressing,Prodromou2012NoCAlert,Parikh2011Formally}\\ \hline \hline
            \multirow{5}{*}{Hard Faults} & \multirow{3}{*}{Routing Technique} & Spare wire \cite{Lehtonen2010Self,Shamshiri2009yield} \\ \cline{3-3}
            &           & Split transmission \cite{Hernandez2008Dealing} \\ \cline{3-3}
            &            & Fault-Tolerant routing algorithm \cite{BenAhmed2013Architecture,DeOrio2012reliable}\\ \cline{2-3}
            & \multirow{2}{*}{Architecture-based Technique} & Hardware Redundancy \cite{Constantinides2006Bulletproof}\\ \cline{3-3}
            &           & Reconfiguration architectures \cite{DeOrio2012reliable} \\ \hline
        \end{tabular}
    \end{center}
\end{table*}        

For soft errors in the data path, most works use code-based techniques that not only detect the integrity of the received data, but also provide a correction function up to an acceptable number of faults. For instance, \textit{Bertozzi et al}.~\cite{Bertozzi2005Error} analyzed several low cost coding techniques for on-chip communication. Among these techniques, \textit{SECDED (Single-Error Correcting and Double-Error Detecting)} was found to be the solution with the most balanced trade-off between reliability and implementation cost. Although the authors provide several evaluations of energy and hardware complexity, on-chip communication analysis (such as throughput and latency) is missing. 
As an adaptive solution, \textit{Yu et al.}~\cite{Yu2010Transient} presented a dynamic \textit{ECC} based on quality of wire connection by using a configurable \textit{ECC} with two Hamming codes to adapt with several probabilities of faults. Although this adaptive \textit{ECC} obtains energy efficiency, its area overhead is problematic.

Soft errors can be detected and recovered using temporal redundancy. For example, {\textit{Ernst et al.}}~\cite{Ernst2003Razor} presented a \textit{Razor D Flip-flop} with an additional shadow latch sampled by a delayed clock for checking the occurrence of transient faults. Furthermore, a soft error detection solution based on redundant latches was also presented by \textit{Ravindan et al.}~\cite{Ravind2009Structural}. Although these techniques obtain more efficient detection results, they nearly double the area overhead and power consumption to maintain the redundant latches. 

For soft errors in the control logic, there are several techniques with cross-layer resolution. 
In the End-to-End level, \textit{Shamshiri et al.}~\cite{Shamshiri2011End} proposed error-correction and on-line diagnosis using a specific code named \textit{2G4L}. Based on the position of the erroneous bit in the received data, the system can indicate the position of the faulty node in the network; however, when a packet is misrouted due to wrong routing information/arbitration or an adaptive routing algorithm, the path of a packet is not fixed in a way that can determine the faulty node. 
To ensure arbitration computation across layers, \textit{NoCAlert}~\cite{Prodromou2012NoCAlert} implements constraints to obtain computational accuracy. 
By constraining the relationship between the input and output of a block, the system can detect both soft and hard faults. 
Although this work presents efficient detection, it lacks efficiency in recovering from soft errors. 
First, the system needs to distinguish between soft and hard faults to decide the recovery method. 
Second, soft errors cannot be recovered by spatial redundancy and their recovery in the End-to-End level is inefficient. The \textit{FoReVer framework}~\cite{Parikh2011Formally} also presented a network level method to detect and recover from routing errors: lost, duplicated, and misrouted packets. 
Since \textit{FoReVer} is based on End-to-End detection and recovery, dealing with soft errors requires retransmission of the whole packet instead of an online recovery. 

In the physical/data-link layers, one of the most common methods is using Triple Modular Redundancy (TMR). 
By triplicating the original module, the system gets three results at the same time~\cite{Radetzki2013Methods}. 
The three results are sent to a \textit{Majority Voting} module to decide the accurate result. 
Although this technique suffers from high area overhead and power consumption (about 300\%), it is easy to implement and effective for both soft errors and hard faults. In \cite{Yu2013Addressing}, the authors deploy a monitoring system on important control modules. They can diagnose the output to find the failure. This technique is light-weight in both area and power and has an insignificant impact on the system performance. However, it suffers from lack of flexibility since the monitor module has to be specifically designed depending on the target component. If any changes in the routing algorithm or pipeline stages are needed, investigation and re-designing of the monitor module is mandatory.

\section{Adaptive 3D Router Architecture (SHER-3DR)} \label{sec:arch}

\begin{figure*}[bht]
    \centering
    \includegraphics[width=1\linewidth]{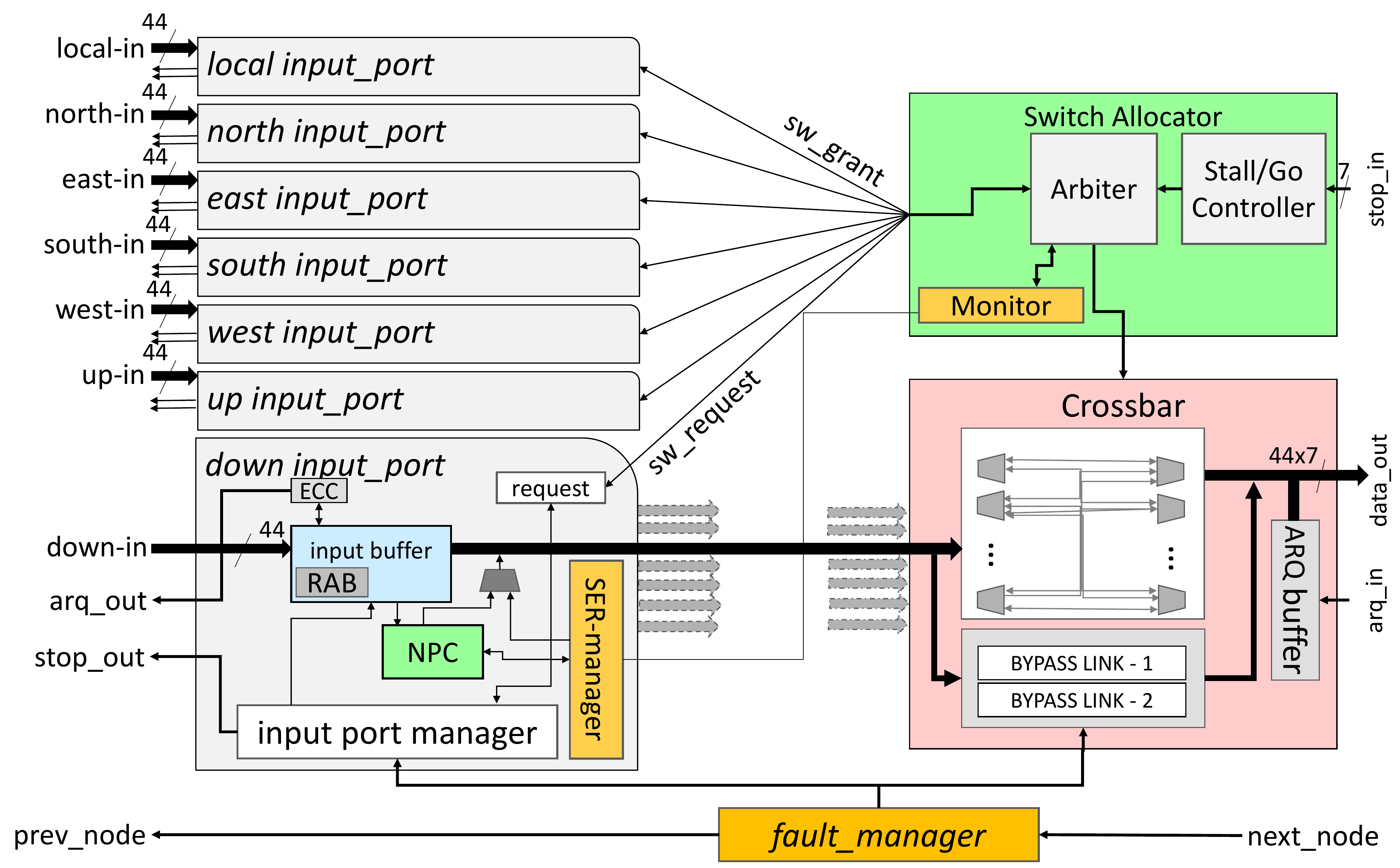}
    \caption{Adaptive 3D router (SHER-3DR) architecture.}
    \label{fig:R_A_2}
\end{figure*}  
Figure~\ref{fig:R_A_2} shows the block diagram of the proposed adaptive 3D router architecture (SHER-3DR). The router relies on simple recovery techniques based on system reconfiguration with redundant structural resources to contain hard faults in the input-buffers, crossbar, and links, in addition to soft errors in the routing pipeline stages. 

The SHER-3DR router is the backbone component of the 3D-FETO system. Each router has a maximum of 7-input and 7-output ports, where 6 input/output ports are dedicated to the connection to the neighboring routers and one input/output port is used to connect the switch to the local computation tile. As shown in {Fig.}~\ref{fig:R_A_2}, the SHER-3DR contains seven \textit{Input-port} modules for each direction in addition to the \textit{Switch-Allocator}, and the \textit{Crossbar} module which handles the transfer of flits to the next node. An \textit{Input-port} module is composed of two main elements: an \textit{Input-buffer }and the \textit{LAFT routing} (Next-Port-Computing) module. Incoming flits from different neighboring routers, or from the connected computation tile, are first stored in the \textit{Input-buffer}. This step is considered to be the first pipeline stage of the flit's life-cycle, Buffer-Writing (BW). After receiving and storing the flits, their routing information is read and processed by a \emph{LAFT-Routing} module (\emph{Next-Port-Computing}) and an arbitrating module (\emph{Switch-Allocator}). This step is the second stage - Next-Port-Computing/Switch-Allocator (NPC/SA). After the NPC/SA pipeline stage, the \emph{next-port} value is merged into the flit and the \emph{grant} signal allows the flit to traverse from its input port to an output port (Crossbar-Traversal (CT) stage).


An augmented Look-Ahead-Fault-Tolerant routing algorithm (LAFT)~\cite{BenAhmed2012LA,BenAhmed2012Low} is used to perform the routing decision. If a given flit is routed to the local port, there is no routing calculation. If the flit is to be routed to another node, the fault link information of all neighboring nodes is read by each input-port and LAFT routing is executed. 
The first phase of the algorithm is calculating the next node's address and its fault output information. In the next phase, the LAFT routing algorithm determines the minimal paths which are valid for routing after eliminating the faulty paths. The final routing path is selected by evaluating two factors of all the possible routing paths: (1) the diversity of the routing path to the destination node and (2) the congestion value of the connection. If there is no minimal routing path, a similar approach is applied for the non-minimal routing paths. Finally, an output port of the selected routing is calculated. This information is merged in the flit as \textit{next-output-port} bits for routing in following nodes~\cite{BenAhmed2016Adaptive}. 

\subsection{Hard Fault Recovery Mechanism Overview}  \label{ssec:HER}
\begin{figure*}[htbp]
    \centering
    \includegraphics[width=1\linewidth]{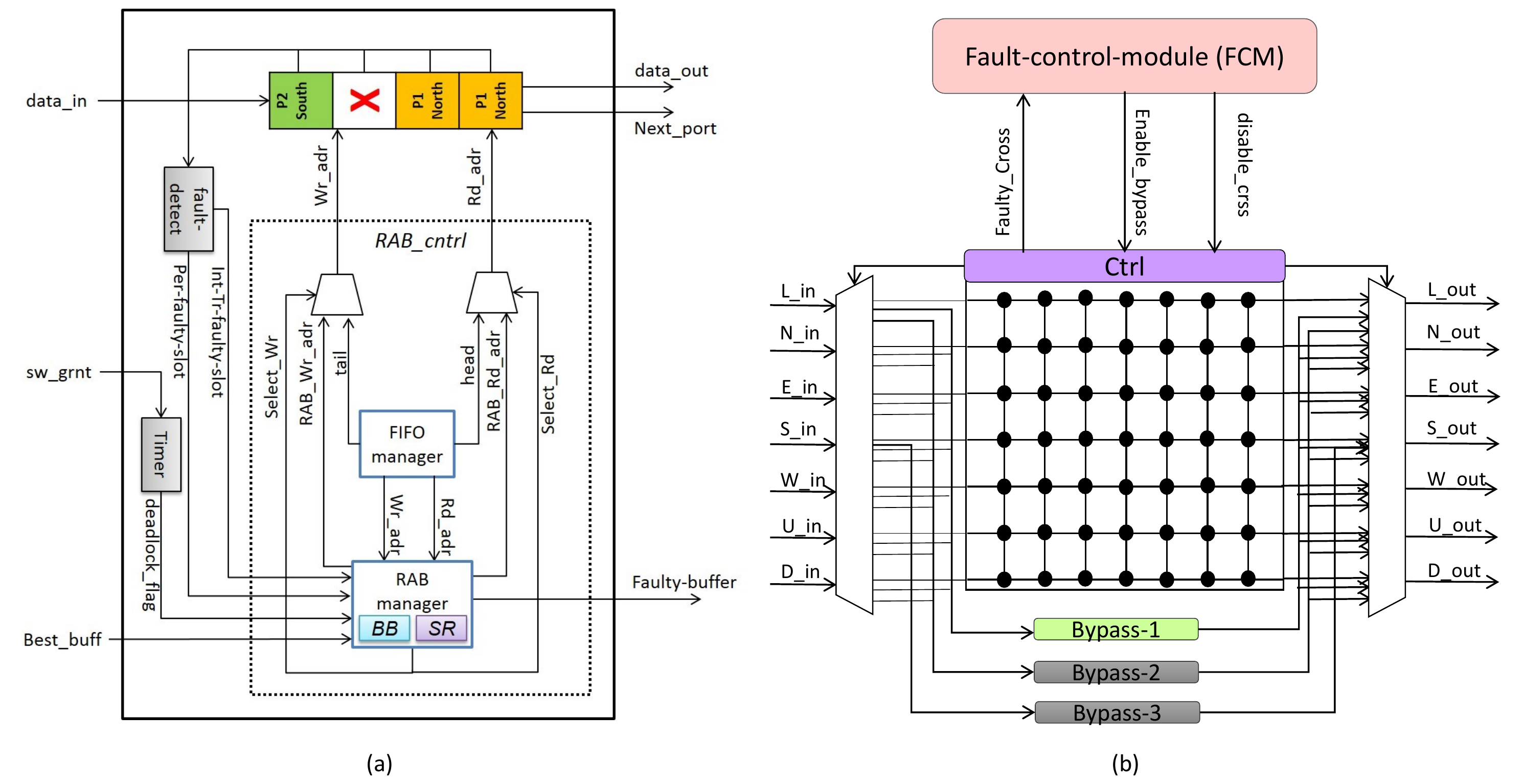}
    \caption{Hard-fault tolerant mechanism~\cite{BenAhmed2016Adaptive}: {(a) Random Access Buffer (RAB); (b) Bypass-Link-on-Demand (BLoD)}}
    \label{fig:3D-FTO-Techniques}
\end{figure*}
%
The block diagram of the hard fault recovery mechanism is shown in Fig.~\ref{fig:3D-FTO-Techniques}. The Random Access Buffer mechanism (RAB) \cite{BenAhmed2016Adaptive} solves the deadlock problem that can occur with the look-ahead fault-tolerant routing algorithm (LAFT), and is able to recover from transient, intermittent, and permanent faults in the input-buffer. When a fault is detected in one of the slots, the main controller (located in \textit{input port manager} in Fig.~\ref{fig:R_A_2})  considers the flagged slots when assigning the write and read addresses. It remains to check the flagged slots for recovery from the faults.

The Bypass Link on Demand mechanism (BLoD)~\cite{BenAhmed2016Adaptive} provides additional escape channels whenever the number of faults in the baseline 7x7 crossbar increases. When a fault is detected in one or several crossbar links, the \textit{fault\_manager} (depicted in Fig.~\ref{fig:R_A_2}) disables the faulty crossbar links and enables the appropriate number of bypass channels.
The number of Bypass-links is very important and it should be minimized as much as possible to reduce the area and power overhead. In a case where the number of faulty links is larger than the number of backup links, the system needs to mark the links as faulty and use the LAFT algorithm to avoid routing through this defective connection.
\begin{figure*}[btph]
    \centering
    \includegraphics[width=.9\linewidth]{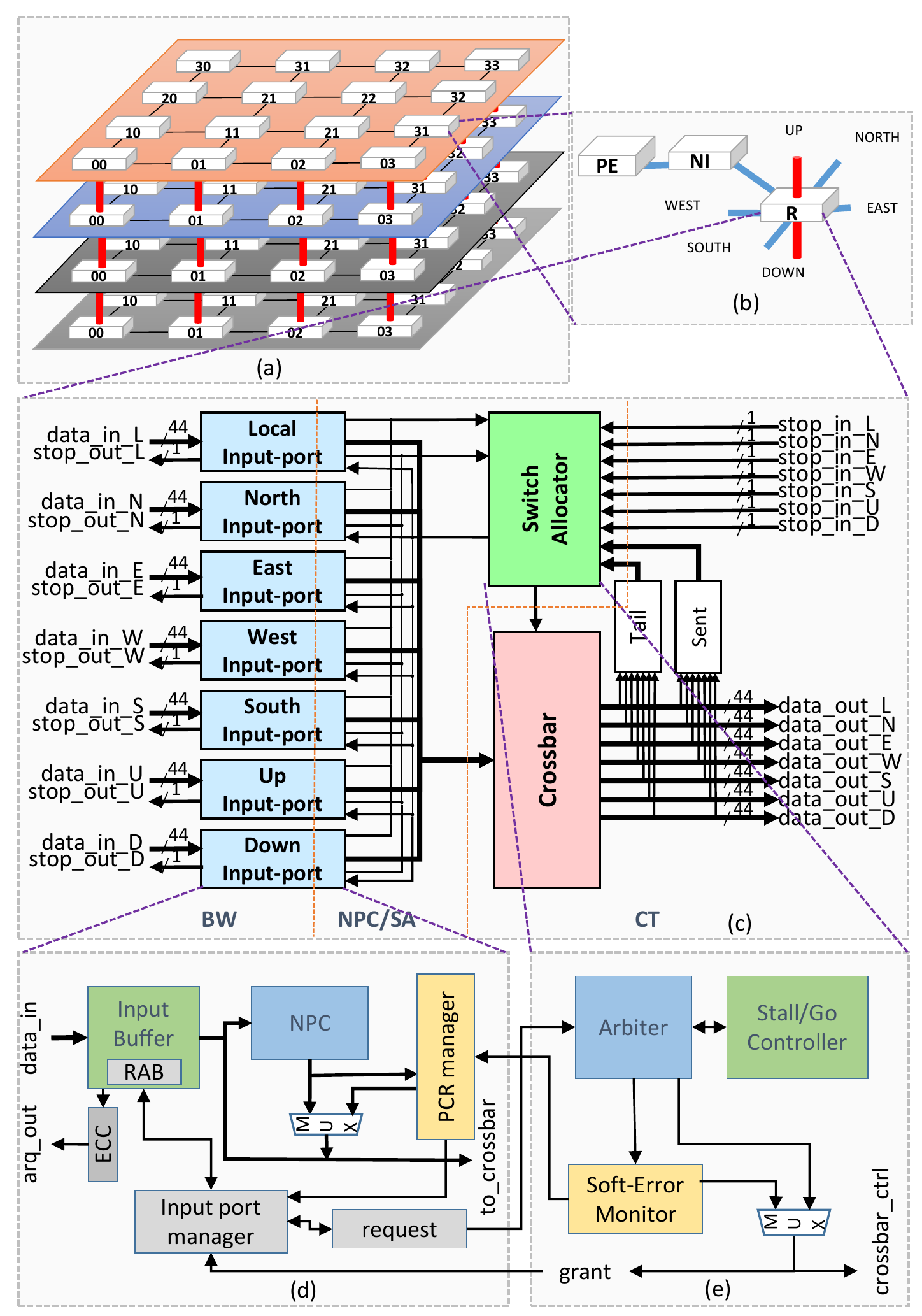}
    \caption{{High-level view of the soft-hard error recovery approach: (a) 3D-Mesh based NoC configuration; (b) Tile organization; (c) SHER-3DR router organization; (d) Input-Port; (e) Switch allocation unit.}}
    \label{fig:SER-3DR-arch}
\end{figure*}  

\subsection{Soft Error Recovery Mechanism}



As represented in Fig.~\ref{fig:SER-3DR-arch}, the principal soft-error handling method in the proposed 3D-FETO system relies 
on a solution called \textit{Pipeline Computation Redundancy} (PCR) in one more clock cycle~\cite{Dang2015Soft}. 

For ease of understanding, we explain {the PCR in Algorithm}~\ref{algo:SERA}. 
{The Next Port Computing (NPC) and Switch Allocator (SA) run in parallel (line 2,3) after the Buffer Writing stage}. This is achieved by the LAFT routing algorithm, where the dependency between the two stages is eliminated. After the first computation, both of the two stages have an additional computation clock cycle (line 4, 5).{ By comparing two consecutive results, soft errors will be detected}. If a soft error is detected, the whole pipeline is halted for correction. {A third computation is required for majority voting, which decides the final result. To recover from soft errors in the data, Single Error Correction Double Error Detection (SECDED)}~\cite{Hsiao1970class} 
{with ARQ (Automatic Retransmission Request)}~\cite{Lin1984Automatic} {is adopted.}
\begin{algorithm}[bhtp]
    \scriptsize
    \caption{{Algorithm of Pipeline Computation Redundancy (PCR)}.}\label{algo:SERA}
    \tcp{input flit's data}
    \KwIn{in\_flit}
    
    \tcp{output flit's data}
    \KwOut{out\_flit}
    
    \vspace*{.2cm}
    
    \tcp{Write flit's data into buffers}
    
    \textbf{BufferWriting}(in\_flit)
    
    \tcp{Compute first time of NPC and SA}
    
    next\_port[1] = \textbf{NextPortComputing}(in\_flit)
    
    grants[1] = \textbf{SwitchAllocation}(in\_flit)
    
    \vspace*{.2cm}
    \tcp{ Compute redundant of NPC and SA}
    
    next\_port[2] = \textbf{NextPortComputing}(in\_flit)
    
    grants[2] = \textbf{SwitchAllocation}(in\_flit)
    
    \vspace*{.2cm}
    \tcp{Compare orginal and redundant to detect soft-error}
    
    \tcp{ Soft-error on NPC}
    
    \uIf {(next\_port[1] $\neq$ next\_port[2])} {
        \tcp{ roll-back and recalculate NPC}
        
        next\_port[3] = \textbf{NextPortComputing}(in\_flit)
        
        final\_next\_port =  \textbf{MajorityVoting}(next\_port[1,2,3]);
    }\Else{
        \tcp{ No soft-error on NPC}
        final\_next\_port = next\_port[1]    
    }
    
    \tcp{Soft-error on SA}
    \uIf {(grants[1] $\neq$ grants[2])}{
        \tcp{roll-back and recalculate SA}
        grants[3] = \textbf{SwitchAllocation}(in\_flit)
        
        final\_grants = \textbf{MajorityVoting}(grants[1,2,3])
    }\Else{ 
        \tcp{No soft-error on SA}
        
        final\_grants = grants[1]
    }
    \tcp{After detection and recovery, the algorithm finishes with CT}
    
    out\_flit = \textbf{CrossbarTraversal}(in\_flit, final\_next\_port, final\_grants);
    
\end{algorithm}

In the first stage, flits are stored in the input buffer at the {Buffer Writing} (BW) stage, and the ECC is used to check and correct the input data in the ECC module. In second stage, the NPC and the SA are executed in parallel in the \textit{LAFT routing} unit and the \textit{Switch-Allocator} module. In third stage, the {Redundant NPC} (RNPC) and the {Redundant SA} (RSA) are computed in parallel. Then, if the output of RNPC is equal to that of NPC, and SA is equal to RSA, the Crossbar Traversal (CT) stage is performed in the third cycle, and the flit goes to the next router via the output channel. If the RNPC is not equal to the NPC, the system rolls-back and recomputes the NPC. Moreover, if SA is not equal to RSA, the system also rolls-back and re-computes the SA stage. After rolling-back and re-computing, a majority voting module is used to decide the correct output of these modules. The rolling-back, re-computing and voting are executed. Then, the outputs of NPC/SA are sent to the {Crossbar Traversal} stage to finish the flit transmission.

\begin{figure}[hbtp]
    \centering
    \includegraphics[width=1\linewidth]{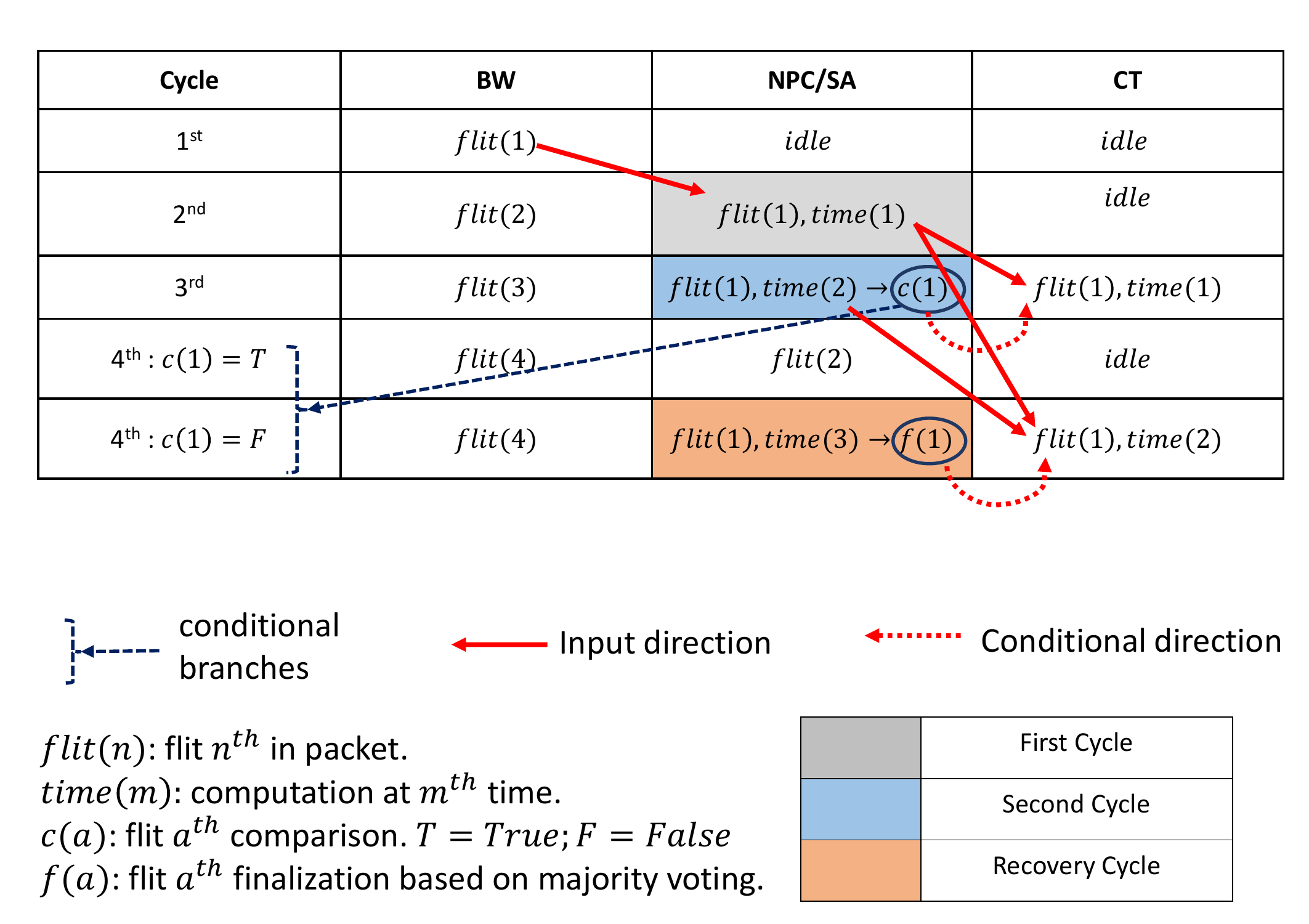}
    \caption{SHER-3DR working demonstration.}
    \label{fig:Transient-Pipeline}
\end{figure}

Figure~\ref{fig:Transient-Pipeline} presents a working demonstration of the SHER-3DR router. $[flit(n)]$ represents the flit in the $n^{th}$ position of the packet. $[time(m)]$ illustrates the $m^{th}$ time of computation. In the first clock cycle, BW handles $[flit(1)]$ while NPC/SA and CT are idle or are handling another packet. In the second cycle, NPC/SA computes $[flit(1), time(1)]$, which means the computation of the first flit for the first time. In the third cycle, NPC/SA computes $[flit(1), time (2)]$, which means that it computes the first flit for the second time, also known as the redundant computation. $[c(1)]$ compares the results of $[flit(1), time(1)]$ and  $[flit(1), time(2)]$ to detect the occurrence of a soft error. If there is no error, CT processes $[flit(1), time(1)]$ to finish the pipeline stages of the first flit. If there is an error in NPC/SA, the system requires the recovery in the fourth cycle. In this cycle, NPC/SA recalculates the first flit for the third time for recovery ($[flit(1), time(3)]$) and finalizes an accurate result by using majority voting ($[f(1)]$). After getting the final result of the first flit, CT completes the pipeline stage of the first flit based on the correct result of the two previous computations: $[flit(1), time(1)]$ or $[flit(1), time(2)]$. As shown in Fig.~\ref{fig:Transient-Pipeline}, the router requires one clock cycle for detecting a soft-error and one optional cycle for recovering each time an error occurs. 

\section{Light-weight Detection, Diagnosis and Recovery Mechanism (DDRM)} \label{ssec:DDRM}


\begin{algorithm}[hbpt]
    \caption{Fault Detection, Diagnosis and Recovery.}
    \label{alg:DRA}
    \tcp{Automatic Retransmission Request}
    \KwIn{$transmitting\_flit$}
    \tcp{Transmitted Buffer Position}
    \KwIn{$buffer\_position$}
    
    \tcp{Control signal to all Fault-Tolerance modules}
    \KwOut{$RAB\_control$, $BLoD\_control$, $LAFT\_control$}
    
    
    \tcp{Transmit the flit, get the ECC's feedback}
    \textbf{Transmit}($transmitting\_flit$);
    
    $ECC\_result$ = \textbf{ECC-Decoder}($transmitting\_flit$);
    
    \tcp{DETECTION PHASE:}
    
    \uIf{$ECC\_result == ARQ$}{
        \tcp{Automatic Retransmission Request}
        \textbf{increase}($ARQ\_counter$);
        
        \textbf{ARQ}($transmitting\_flit$);
    } \Else {
    \tcp{The transmitted flit is non faulty}
    \textbf{Finish};
}
\tcp{Check the number of consecutive ARQs}
\If{($ARQ\_counter == 2$)}{
    \tcp{There is a permanent fault}
    \tcp{Jump to DIAGNOSIS-RECOVERY PHASE}
}

\tcp{DIAGNOSIS-RECOVERY PHASE:}
\tcp{Start with Input Buffer Checking}
$Buffer\_Failure \leftarrow Buffer\_Checking(buffer\_position)$;

\uIf{($Buffer\_Failure == Yes$)}{
    \tcp{Random Access Buffer is received the position to handle.}
    
    $RAB\_Control = buffer\_position$;
    
    \textbf{Finish};
}\Else{ 
\tcp{The buffer slot is non faulty.}
\tcp{Move to Crossbar Checking: using a Bypass-Link.}
$BLoD\_control$ = enable;

\tcp{Get the ECC's feedback and detect with ARQ counter.}

\uIf{($ARQ\_counter == 2$)}{
    \tcp{BLoD cannot fix the fault, the link is failed.}
    
    $BLoD\_control$ = release;
    
    \tcp{The LAFT routing algorithm handles the faulty link.}
    $LAFT\_control$ = faulty;
    
    \textbf{Finish};
}\Else{
\tcp{BLoD already fixed the failure, the recovery step is finished.}
\textbf{Finish};
}

}
\end{algorithm}

Algorithm~\ref{alg:DRA} shows the proposed \textit{Detection, Diagnosis and Recovery Mechanism} (DDRM). It uses the feedback from the ECC and the Automatic Retransmission Request (ARQ) protocol to monitor the errors. 
As shown in Fig.~\ref{fig:R_A_2}, the input data is first verified by an ECC decoder. If the value is correct or the ECC decoder can handle the correction, the flit is  written to the input buffer. Otherwise, a retransmission is requested. Since the transient fault only occurs over a short period of time, assumed to be a single clock cycle, it does not occur for two consecutive cycles. Therefore, ARQ  can recover this kind of fault. However, if a permanent fault occurs, ARQ is unable to correct it and the faulty connection will keep requesting retransmission infinitely. Therefore, if the ARQ cannot correct the fault, the system considers it to be a permanent fault (line 1-10 in Algorithm~\ref{alg:DRA}).

\begin{figure*}[htp]
    \centering
    \includegraphics[width=1\linewidth]{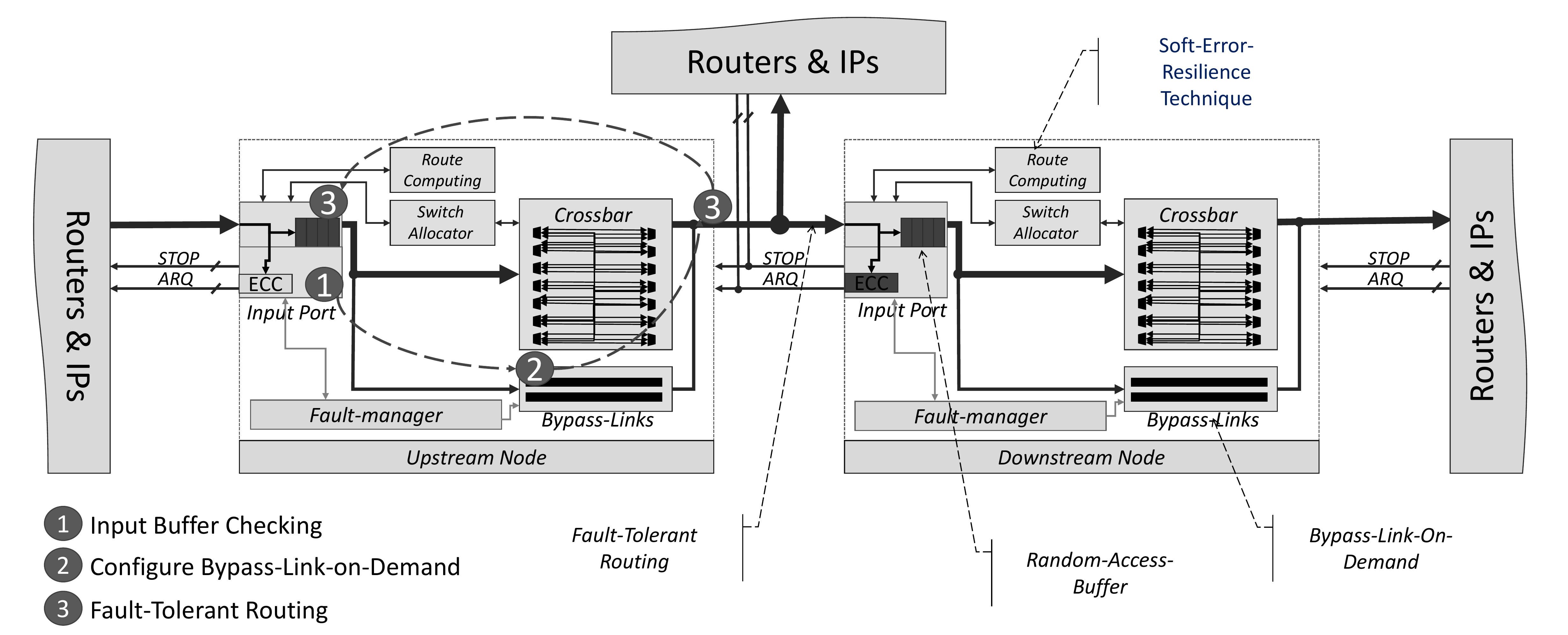}
    \caption{Router-to-Router interfacing and DDRM scheme.}
    \label{fig:r2r}
\end{figure*}

Since a flit's correctness is verified by the ECC module before being written to the buffer, a permanent fault can only occur in the path between the input-buffer in the upstream node and the one in the downstream node. Figure~\ref{fig:r2r} shows the high-level view of the DDRM and Router-to-Router interfacing. The transmission path of a flit consists of 3 main components: input buffer slots, a crossbar link and a router-to-router channel. When a fault is detected, \emph{DDRM} diagnoses these two components to find the fault position and recover it with an appropriate mechanism.

For the diagnosis and recovery phase, the router's \textit{Fault-manager} module initiates the diagnosis with input buffer checking. In this step, the error statuses of the following flits of the monitored input buffer are checked. If errors are detected in the following flits' transmission, it means the fault should belong to the crossbar link or the inter-router channel. The diagnosis is forwarded to check the crossbar and inter-router channel. If errors are constantly detected at the same position of the monitored buffer, the fault belongs to this detected position.
In this fashion, the \textit{Fault-manager} sends a signal to the \textit{Random Access Buffer} (RAB) mechanism to indicate the faultiness of the slot in the input buffer (line 11-14). 
If the fault-manager indicates that the fault may belong to the crossbar or inter-router channel, the \textit{Fault-manager} first configures the \textit{Bypass-Link-on-Demand} (previously presented in Section~\ref{ssec:HER}) to establish an alternative connection path. Then, another flit is sent from the input buffer through a bypass-link and the router-to-router channel to the downstream node. If, at the downstream node, the flit is found  to be not faulty by the ECC module, the \textit{Fault-manager} concludes that the fault is in the Crossbar, which is already handled by the BLoD mechanism. Therefore, the configuration of the BLoD is kept as a recovery.
If the flit is still faulty, the fault belongs to the inter-router channel. In this situation, the BLoD is released for further fault-tolerance and the information of the faulty channel is sent to the routing module (in LAFT algorithm). At the routing module, the \textit{Look-Ahead Fault-Tolerant} routing algorithm uses the fault information to handle the channel's failure. The flit in the input buffer is re-routed via an alternative output port. 
\section{Evaluation Results}\label{sec:eva}

\subsection{Evaluation Methodology}\label{sec:meth}
The proposed 3D-FETO system was designed in Verilog-HDL, synthesized and prototyped with commercial CAD tools and VLSI technology, respectively \cite{NCSUEDA2015FreePDK3D45,nangate2014nangate}. We evaluate the hardware complexity of the SHER-3DR router in terms of area utilization, power consumption (static and dynamic), and speed. To evaluate the performance of the proposed system, we select both synthetic and realistic traffic patterns as benchmarks. For synthetic benchmarks, we selected Transpose \cite{Chien1995Planar}, Uniform \cite{Sivaram1992Queuing}, Matrix-multiplication \cite{Chen2010parallel,Zekri2006general}, and Hotspot 10\%~\cite{Dally2004Principles}. For realistic benchmarks, we chose  H.264 video encoding system~\cite{Rahmani2014High}, Video Object Plane Decoder (VOPD), Picture In Picture (PIP) and Multiple Window Display (MWD)~\cite{Bertozzi2005NoC}. The simulation configurations are depicted in Table~\ref{tab:sim-conf}.

\begin{table*}[htbp]
    \begin{center}
        \caption{Simulation configurations.}
        \label{tab:sim-conf}
        \begin{tabular}{|c|c||c|}
            \hline
            \multicolumn{2}{|c||}{\textbf{Parameter/System}} & \textbf{Value}  \\ \hline \hline
            \multirow{8}{*}{Network Size ($x\times y\times z$)} & Matrix & $6\times6\times3$  \\
            & Transpose & $4\times4\times4$ \\
            & Uniform                 & $4\times4\times4$ \\
            & Hotspot 10\%            & $4\times4\times4$ \\
            & {H.264}                   & $3\times3\times3$ \\   
            & {VOPD}                    & $3\times2\times2$ \\
            & MWD                     & $2\times2\times3$ \\
            & PIP                     & $2\times2\times2$ \\     \hline
            \multirow{8}{*}{Total Injected Packets} & Matrix       	& 1,080  	\\
            & Transpose     	& 640  		\\
            & Uniform      	& 8,192  	\\
            & Hotspot 10\% 	& 8,192  	\\ 
            & {H.264}     		& 8,400     \\
            & {VOPD}     		& 3,494     \\
            & MWD      		& 1,120     \\
            & PIP      		& 512       \\    \hline
            
            \multirow{2}{*}{Packet's Size}  & Hotspot 10\%  &  10 flits + 10\% for hotspot nodes  \\ 
            & Others   &  10 flits   \\ \hline
            \multicolumn{2}{|c||}{Flits Size}    & 44 bits     \\ \hline
            \multicolumn{2}{|c||}{Header Size}   & 14 bits     \\ \hline
            \multirow{2}{*}{Payload Bit}  & Baseline, 3D-FTO  & 30 bits   \\ 
            & Soft Error Tolerance, 3D-FETO      &  18 bits  \\ \hline
            \multirow{2}{*}{Parity Bit}   & Baseline, 3D-FTO  & 0 bits    \\ 
            & Soft Error Tolerance, 3D-FETO      &  12 bits ($2 \times$ SECDED(22,16)) \\ \hline
            
            \multicolumn{2}{|c||}{Buffer Depth}       & 4               \\ \hline
            \multicolumn{2}{|c||}{Switching}          & Wormhole-like   \\ \hline
            \multicolumn{2}{|c||}{Flow-control}      & Stop-Go         \\ \hline
            \multicolumn{2}{|c||}{Routing}            & LAFT           \\ \hline
        \end{tabular}
    \end{center}
\end{table*}        

The above synthetic benchmarks help us understand the performance of the network under stress; however, we also need several realistic benchmarks to understand the network under real application traffic. Therefore, we build a simulator in Verilog-HDL which allows us to set up the traffic patterns from real applications. Based on the traffic patterns, the \textit{Network Interfaces} send and receive packets over the networks. We select a video encoding system using a H.264 encoder, a MP3 encoder, and a OFDM ~\cite{Rahmani2014High}.
%
Moreover, we select three applications~\cite{Bertozzi2005NoC}: VOPD, PIP and MWD.

We evaluate the performance of our fault-tolerant model which includes hard fault tolerance from 3D-FTO~\cite{BenAhmed2016Adaptive}, Soft-Error Tolerance OASIS system, and the proposed system (3D-FETO). We measure the average packet latency, with the selected synthetic and realistic benchmarks. To understand the impact of fault-tolerance techniques on performance, we compare the obtained results with the baseline 3D-NoC system presented in \cite{BenAhmed2012LA}. We randomly inject faults at three fault-rates: 10\%, 20\% and 33\%. The faults are injected into hard fault tolerant and soft error tolerant modules. For the soft error tolerant system, only soft errors are injected. For the hard fault tolerant (3D-FTO) system, only hard faults are injected. For the final system (3D-FETO), both soft errors and hard faults are injected. {Hard faults are injected at the beginning of simulation and their rate is measured as the percentage of routers with faults. Soft errors are injected during the system's operation and their rate is considered to be the number of soft errors per clock cycle. The injected fault rates are considered individually for each error type. }

\subsection{Complexity Evaluation}
\begin{figure}
\centering
\includegraphics[width=0.4\linewidth]{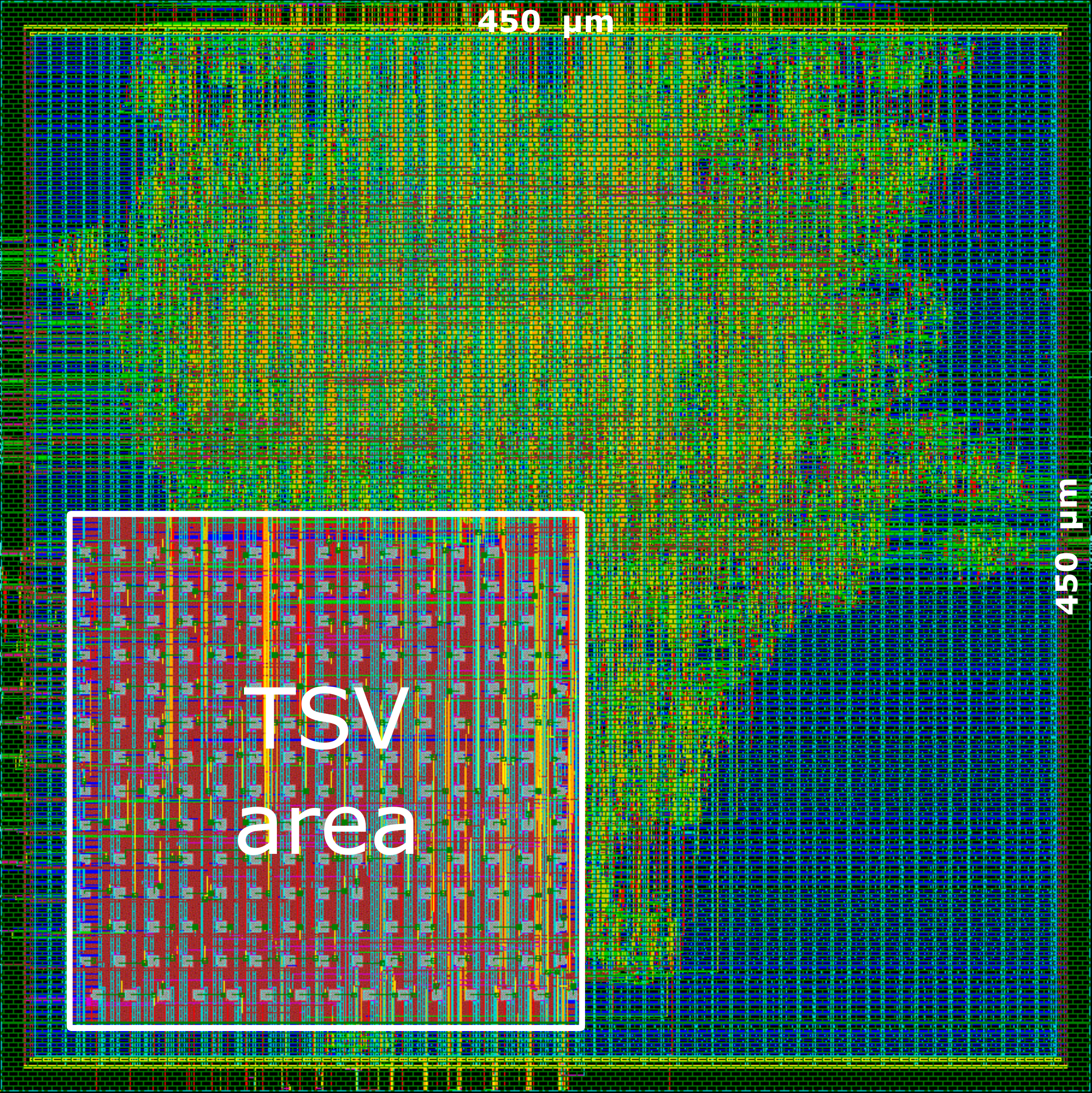}
\caption{Layout of a single SHER-3DR router for the 3D-FETO system. The SHER-3DR router was designed in Verilog-HDL and synthesized using 45nm technology library~\cite{nangate2014nangate}. For the {Through Silicon Via (TSV)} integration, we used FreePDK3D45 kit compiler~\cite{NCSUEDA2015FreePDK3D45}. The SHER-3DR router is designed on a $450\mu m \times 450\mu m$ and the TSV array is 208 TSVs. }
\label{fig:layout}
\end{figure}

\begin{table*}[htbp]
    \begin{center}
        \caption{Hardware complexity evaluation and comparison results.}
        \label{tb:fto-eval}
        \begin{tabular}{|l||c|c|c|c|c|}
            \hline
            & \textbf{Area} & \multicolumn{3}{c|}{\textbf{Power}} & \textbf{Speed}  \\
            \textbf{Model}             & ($\mu m^2$)   & \multicolumn{3}{c|}{(mW)}           & (Mhz)           \\ \cline{3-5}
            &               & Static   & Dynamic & Total          &                 \\ \hline \hline
            Baseline LAFT router         & 18,873        &  5.1229  & 0.9429  & 6.0658         & 925.28          \\
            3D-FTO router                & 19,143        &  6.4280  & 1.1939  & 7.6219         & 909.09          \\
            Soft Error Tolerance router & 27,457        &  9.7314  & 2.6710  & 12.4024        & 625.00          \\
            SHER-3DR               & 29,516        &  10.0819 & 2.7839  & 12.8658        & 613.50          \\ \hline
        \end{tabular}
    \end{center}
\end{table*} 
In this evaluation, we considered the hardware complexity of the proposed SHER-3DR router. {For this evaluation, we use the NANGATE 45$nm$ technology library}~\cite{nangate2014nangate}. {Area cost and power consumption analyses are performed with the Synopsys $\copyright$ Design Compiler. The power consumption information is analyzed based on the switching activity of the router under the uniform benchmark.} We start first by observing the additional hardware added to the baseline system when we employ the hard fault tolerance model (3D-FTO router). Then, we evaluate the impact when we consider the soft error tolerant model (Soft Error Tolerant router). Finally, we evaluate the completed SHER-3DR system including both soft and hard fault tolerant mechanisms. The configurations of the network are shown in Table~\ref{tab:sim-conf} and the layout of a single SHER-3DR router is depicted in Fig.~\ref{fig:layout}.

Table~\ref{tb:fto-eval} illustrates the hardware complexity results of SHER-3DR router in terms of area, power (static, dynamic, and total), and speed. In the hard fault tolerance router (3D-FTO), the area and power consumption overheads have increased by {1.43\%} and 25.65\%, respectively. The maximum speed has also slightly decreased. On the other hand, our soft error handling mechanism adds seven ARQ buffers and some combinational logic which increase the area and power consumption more significantly. However, SHER-3DR introduces 7.50\% and {3.74\%} extra area and power consumption, respectively, when compared to the soft error tolerant model. In comparison to the baseline model, SHER-3DR increases the area and power consumption by 56.39\% and 112.10\%, respectively, while the maximum speed decreases by 33.70\%.

\begin{figure}
    \centering
    \includegraphics[width=1\linewidth]{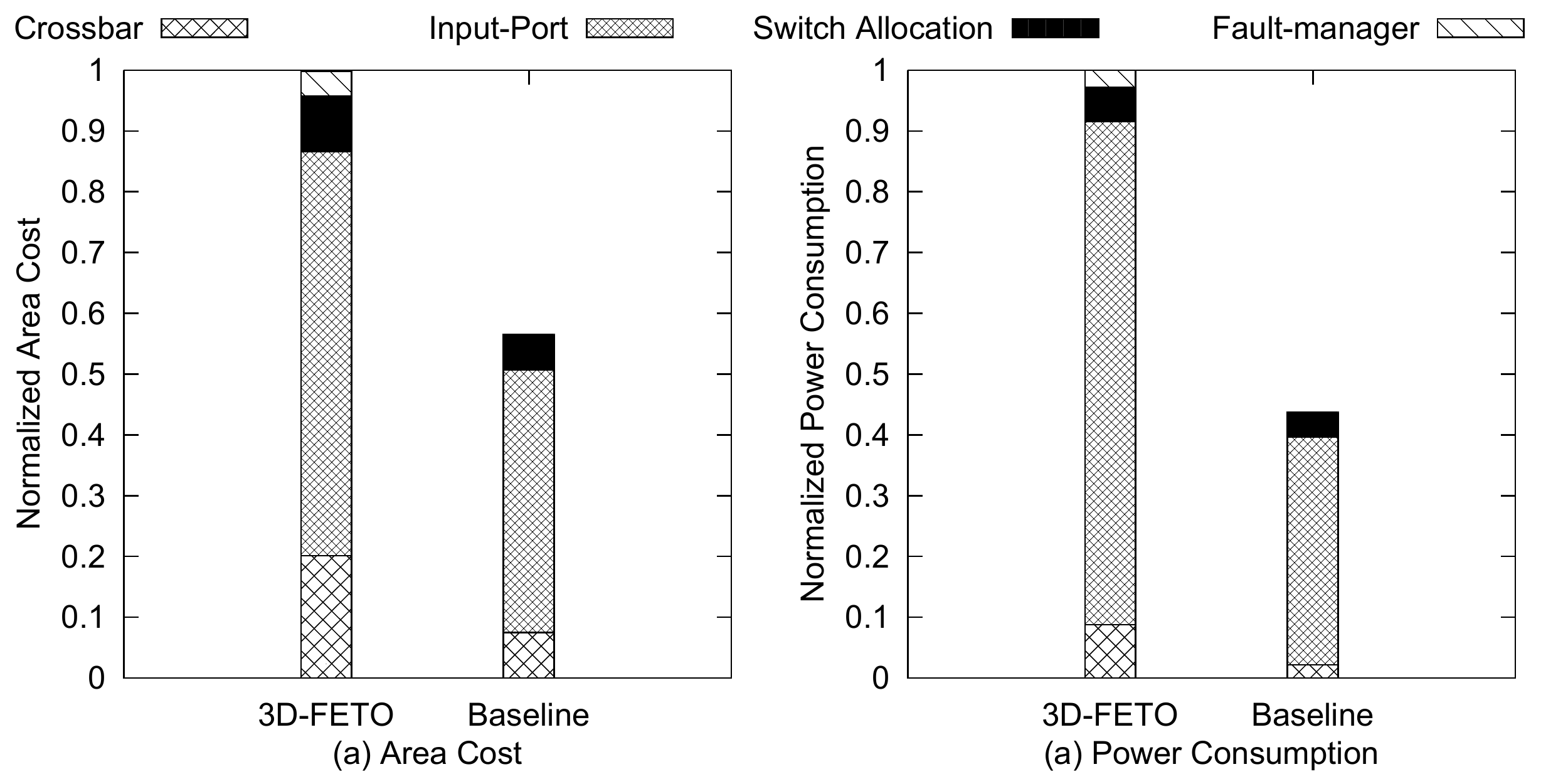}
    \caption{{Area cost and power consumption analysis.}}
    \label{fig:complex-ana}
\end{figure}

{The area cost and power consumption of the proposed router is given by Equation}~\ref{eq:area} {where $\pi_i$ represents the area cost or power consumption of module $i$. 
The SHER-3DR router consists of four main modules: input-ports, switch-allocator, crossbar, and fault manager.} 
\begin{equation} \label{eq:area}
\pi_{router} = \pi_{input-ports} + \pi_{switch-allocator} + \pi_{crossbar} + \pi_{fault-manager} 
\end{equation}
{The details of an input port, a switch-allocator and a crossbar are given in Equation}~\ref{eq:modulearea}. 
\begin{multline} \label{eq:modulearea}
\pi_{input-ports} = \pi_{original-input-ports} +  \pi_{RAB-controller} +  \pi_{PCR-controller} +  \pi_{ECC} \\
\pi_{switch-allocator} = \pi_{original-switch-allocator} + \pi_{PCR-monitor} \\
\pi_{crossbar} = \pi_{original-crossbar} +  \pi_{bypass-links} +  \pi_{ARQ-buffers} \\
\end{multline}
{We can observe the overheads in power consumption and area cost that are caused by the fault-tolerance mechanisms (RAB-controller, PCR-controller, ECC, BLoD, ARQ buffers). Figure} \ref{fig:complex-ana} {provides the evaluation results of power consumption and area cost of SHER-3DR. In terms of area cost, the input ports occupy the majority with over 67\% which is followed by the crossbar (20\%) and the switch allocator (9\%). The fault manager, which supports DDRM, uses only about 4\% of the overall area cost. In terms of power consumption, the input ports consume over 80\% of the total value. The fault manager module also causes an insignificant increase in power consumption (3\%).} 

{When compared to the baseline OASIS router, the proposed SHER-3DR consumes more power consumption and costs more area. As shown in Fig.}~\ref{fig:complex-ana}, {SHER-3DER increases the area and power of all three main modules (crossbar, input ports, and switch-allocator). The overhead can be analyzed by Equation}~\ref{eq:modulearea}{ where additional modules are attached to support the fault-tolerance mechanisms.}

Although our proposed models are penalized in terms of area, power consumption, and maximum frequency due to additional logic and registers that are necessary for fault handling mechanisms, they provide an improved resiliency against a significant amount of soft and hard faults.

\subsection{Latency Evaluation}
\begin{figure*}[bht]
    \begin{center}
        \includegraphics[width=.9\linewidth]{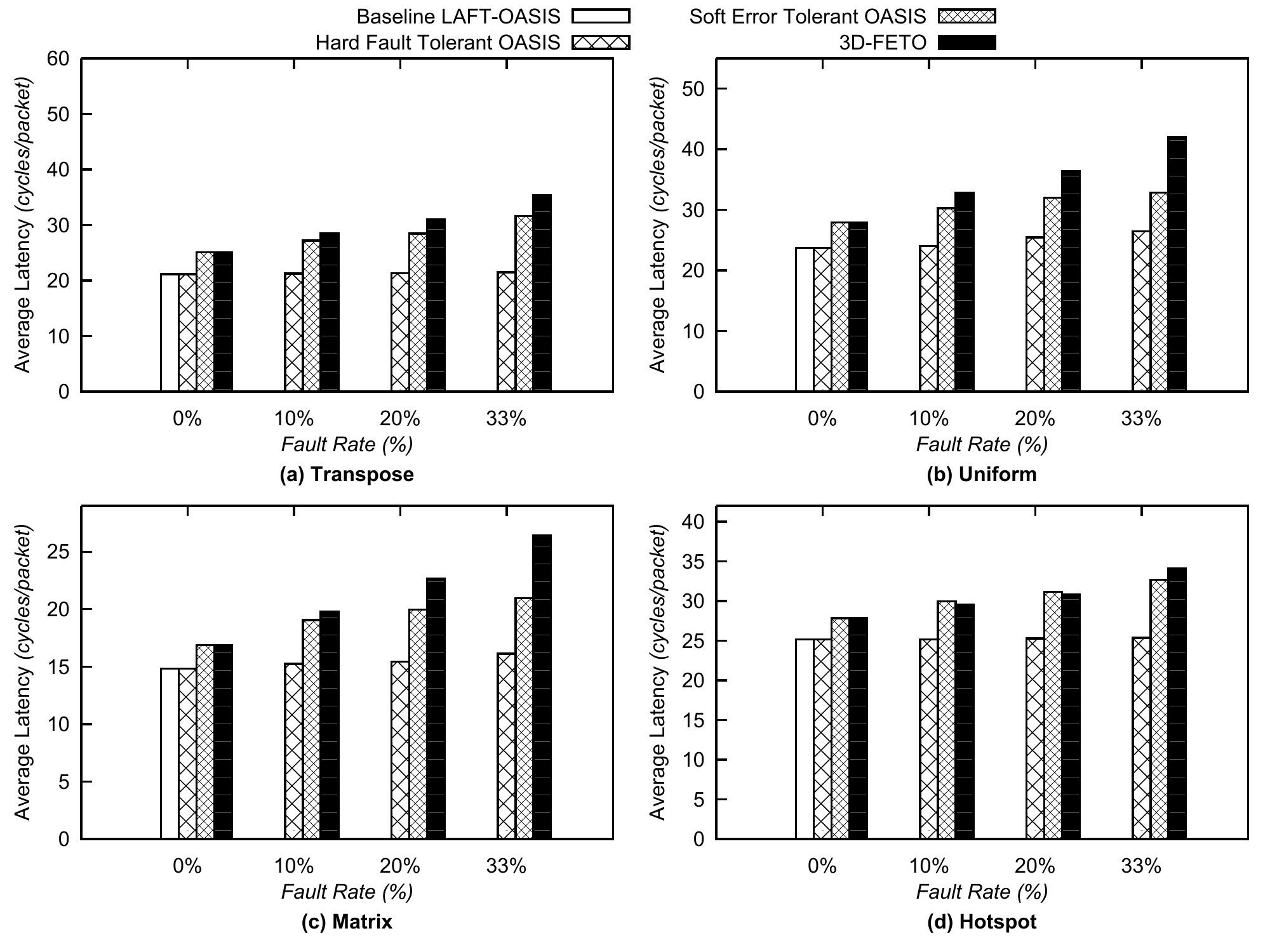}
        \caption{Average packet latency evaluation of the synthetic benchmarks.}
        \label{fig:syn-lat}
    \end{center}
\end{figure*}

\begin{figure*}[bht]
    \begin{center}
        \includegraphics[width=.9\linewidth]{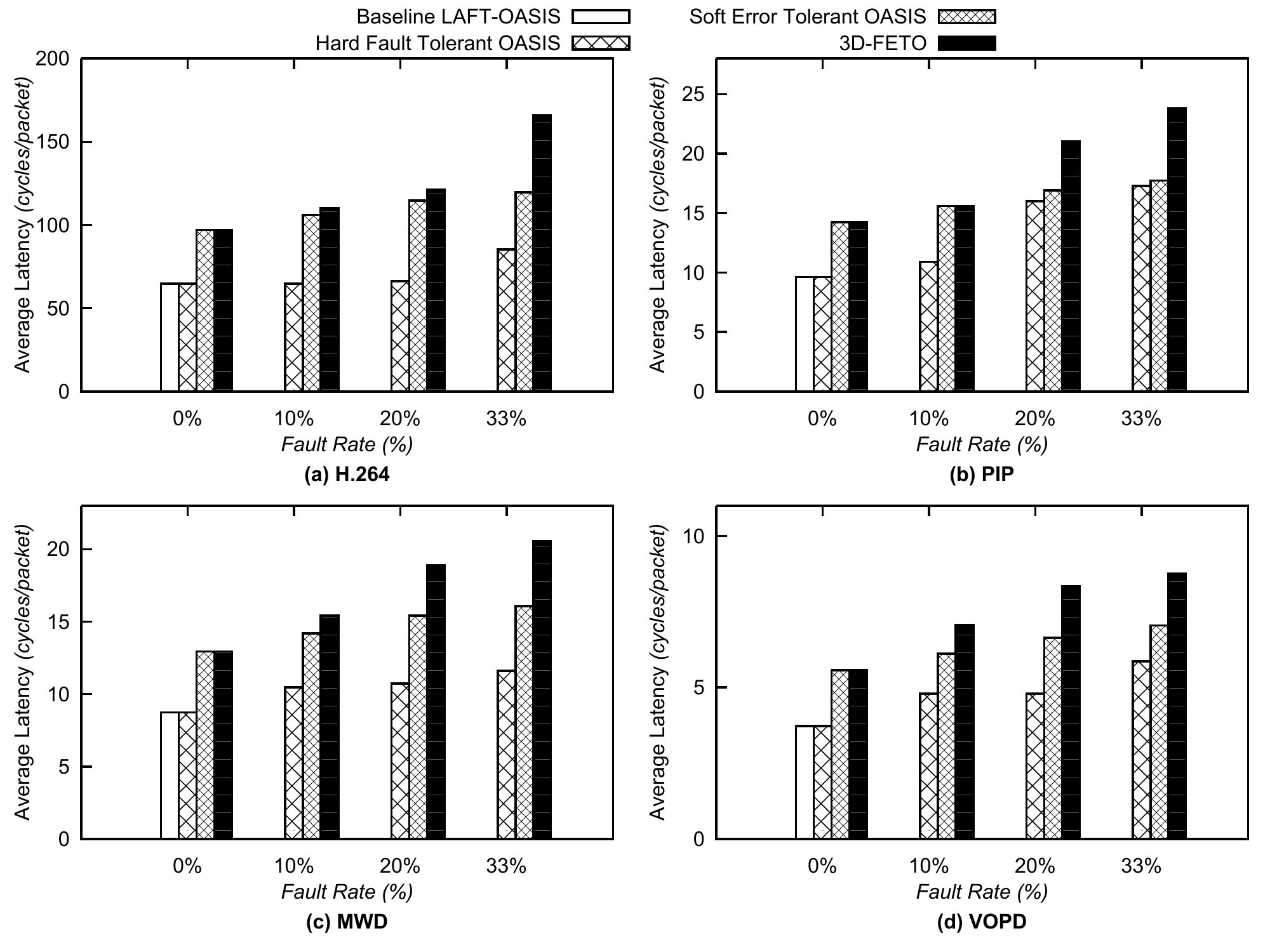}
        \caption{Average packet latency evaluation of the realistic benchmarks.}
        \label{fig:real-lat}
    \end{center}
\end{figure*}

In the second experiment, we evaluate the performance of the proposed architecture in terms of latency over various benchmark programs and error injection rates for three system configurations: (1) Hard-fault tolerant system (3D-FTO), (2) Soft-error tolerant OASIS system, and (3) Hard-fault and Soft-error tolerant system (3D-FETO). The simulation results are shown in Figs.~\ref{fig:syn-lat} and ~\ref{fig:real-lat}. From these graphs, we notice that with 0\% hard faults (in input buffer and crossbar only), 3D-FTO has similar performance to the baseline system (LAFT-OASIS). In addition, we found that even at a 33\% fault-rate, 3D-FTO increases the latency by only 1.71\%, 11.38\%, 8.79\% and 13.73\% for Transpose, Uniform, $6\times 6$ Matrix, and Hotspot-10\%, respectively. 
With realistic benchmarks, the performance of 3D-FTO slightly degrades at low error-rates, but it suffers more of an impact at high error-rates (20\% and 33\%) since the flit encounter bottlenecks due to errors inside the input buffers. However, the proposed 3D-FETO model still works even at high fault-rates while the baseline model collapses at a 5\% error-rate. 
We used the same benchmark programs to evaluate the soft error tolerant model. Since both the proposed \textit{Pipeline Computation Redundancy} mechanism and ECC require additional clock cycles, we can observe a significant effect on average packet latency. For the 0\%, 10\%, 20\% and 33\% fault-rates, the 
Soft Error Tolerant model increases the average delay in the Transpose benchmark by 18.57\%, 28.74\%, 34.54\% and 49.62\%, respectively.  
Finally, we evaluate the proposed 3D-FETO system with both soft error and hard fault handling schemes. As shown in Figs.~\ref{fig:syn-lat} and~\ref{fig:real-lat}, 3D-FETO has demonstrated a significant impact on the average latency, which has mostly doubled for both realistic and synthetic benchmarks. At a 33\% fault-rate using Matrix, Uniform, Transpose benchmarks, 3D-FETO's average latency increases by 78.44\%, 50.73\% and 67.18\% in terms of average packet latency. {The degradation is caused by both soft errors and hard fault tolerance mechanisms: (1 the) ECC+ARQ and PCR both require additional re-transmission clock cycles; (2) the RAB and LAFT routing algorithm may disable a part of the network which causes congestion.} However, it still maintains the ability to work under an extremely high fault-rate (33\% for hard faults and 33\% for soft errors).
%
%

%
%
\subsection{Throughput Evaluation}
%
\begin{figure*}[bht]
    \begin{center}
        \includegraphics[width=.9\linewidth]{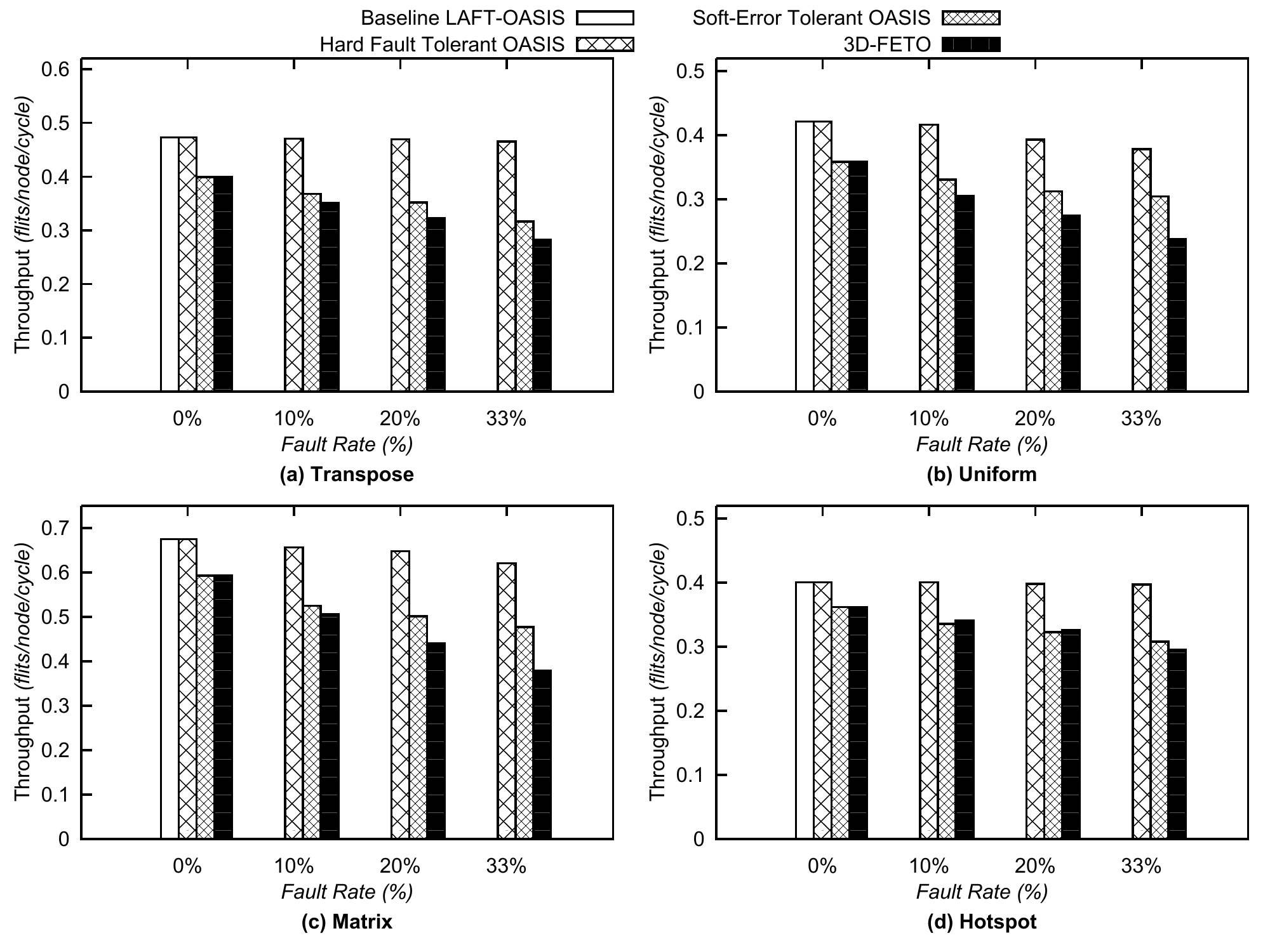}
        \caption{Throughput evaluation of the synthetic benchmarks.}
        \label{fig:syn-tp}
    \end{center}
\end{figure*} 
%
Figure~\ref{fig:syn-tp} depicts the throughput evaluation with the adopted synthetic benchmarks. At a 0\% error rate, 3D-FTO (hard-fault tolerance) presents the best throughput which matches the capacity of the baseline LAFT-OASIS. The Soft Error Tolerant OASIS and the proposed 3D-FETO have less throughput due to their soft error tolerance mechanisms. When the errors are injected into the system, we can observe a degradation in throughput. Thanks to the efficient hard fault tolerance scheme and the fault-tolerant routing algorithm, 3D-FTO at a 33\% error-rate provides a slightly decreased throughput: 40.18\%, 43.96\%, 43.55\% and 32.59\% for Transpose, Matrix, Uniform, and Hotspot 10\%, respectively. 
For the Soft Error Tolerant OASIS, the system requires re-transmission via the ARQ mechanism and the re-execution for the soft error mechanism. Therefore, the throughput is degraded due to extra clock cycles. {The proposed 3D-FETO, which is a fusion of both hard fault tolerance and soft error tolerant mechanisms, inherits both degradations}; however, these systems provide the ability to handle up to a 33\% error rate (the limitation of the soft error mechanism).

\begin{table}[htbp]
    \begin{center}
        \caption{Successful arrival-rate comparison results for a {$5\times5\times4$} system configuration under Uniform traffic.} 
        \label{tb:arrival-uni}
        \begin{tabular}{|l|c|c|c|c|c|}
            \hline 
            Algorithm / Fault-rate&1\%&5\%&10\%&15\%&20\%\\ 
            \hline \hline 
            XYZ&91\%&62\%&41\%&28\%&23\%\\ 
            \hline
            Hybrid-XYZ&99\%&83\%&62\%&44\%&36\%\\ 
            \hline
            8-RW&100\%&95\%&85\%&69\%&59\%\\ 
            \hline
            Odd-Even&96\%&85\%&67\%&53\%&43\%\\ 
            \hline
            Hybrid-Odd-Even&100\%&94\%&83\%&70\%&61\%\\ 
            \hline
            4N-FIRST&98\%&89\%&72\%&68\%&46\%\\ 
            \hline
            4NP-FIRST&97\%&98\%&95\%&83\%&76\%\\ 
            \hline
            LAFT-OASIS &100\%&100\%&99\%&98\%&95\%\\ 
            \hline
            3D-FETO& 100\% & 100\% & 99\% & 99\% & 97\% \\ 
            \hline
        \end{tabular} 
    \end{center}
\end{table}
%
\subsection{Reliability Evaluation}
\subsubsection{Arrival Rate}
This subsection presents the reliability evaluation of the proposed 3D-FETO system over several hard fault and soft errors injection rates. For comparison, seven systems adopting different routing algorithms are selected~\cite{Pasricha2011low}: XYZ, Hybrid-XYZ, 8-Random-Walk (8-RW), Odd-Even, Hybrid-Odd-Even, 4N-First, and 4NPFirst. Among these algorithms, we can find deterministic 3D routing algorithms, fault-tolerant 2D algorithms that were extended to the third dimension, and also turn-model based schemes that were proposed for fault-tolerant 3D-NoC systems. We adopted the same simulation environment and assumptions made in~\cite{Pasricha2011low} from where the arrival-rate results were also obtained. For fair comparison, we assume that the faults can occur at any link with LAFT; thus, we eliminate the two assumptions that are necessary for the algorithm to efficiently work: (1) the links connecting the PE to the local input and output ports are always non-faulty. (2) There exists at least one non-faulty path between a (source, destination) pair.  Moreover, we also evaluate the arrival-rate of our final system with the enhancements by Random-Access-Buffer and Bypass-Link-on-Demand. Instead of only distributing faults on the inter-router channel, they are randomly assigned to input buffers, crossbar, or the inter-router channel.

Table \ref{tb:arrival-uni} and Table \ref{tb:arrival-tran} depict the arrival-rate results for a $5\times5\times4$ system (100 nodes) under Uniform and Transpose traffic patterns, respectively. Due to its lack of support for fault-tolerance, XYZ routing demonstrates the worst Arrival-rate for both applications. Its variant Hybrid-XYZ shows slightly better results, but it is still considered unacceptable. Despite the fact that 8-RW is fault-tolerant, its Arrival-rate considerably degrades as we increase the fault-rate. This can be explained by the frequent deadlock-occurrence with this algorithm that is considered to be one of its main drawbacks. 4NP-FIRST is a fault-tolerant routing algorithm targeted for 3D-NoCs. However, it does not scale very well when we increase the fault-rate. In fact, one third of the injected packets fail to reach their destinations at a 20\% fault-rate, which can be seen in Table \ref{tb:arrival-tran}.

Among the considered algorithms, 3D-FETO appears to be the most reliable solution, providing a scalable arrival-rate that does not go under 97\% in both applications, even at a 20\% fault-rate. When observing the results with the two applications, 3D-FETO with the LAFT algorithm is considered to be the only scheme that takes advantage of long distance communications in Transpose traffic. This is in contrast with the remaining algorithms where their reliability degrades considerably with this application. In fact, the combination of look-ahead routing and the path prioritization using the diversity value in LAFT significantly increases the probability for packets to find non-faulty paths to reach their destinations.

The arrival rates of the proposed 3D-FETO reach over 97\% in the worst case (20\% fault-rate) while LAFT-OASIS's arrival rates are 95\% and 96\%. With other rates, 3D-FETO presents its capacity for high reliability with an arrival-rate of over 98\%.
When we analyzed the possible causes for the failing 5\%, we observed the occurrence of cases where all the connecting links of a given router are faulty: for example, the East, North, and UP links of the bottom-left router of the network are broken. Thus, the router cannot receive or inject any flit from/to the network. Another failure case manifests when the link connecting the router to the attached PE is faulty. As expected, these two cases justify the two assumptions that we previously made to ensure the efficiency of LAFT's fault-tolerance capabilities. 

\begin{table}[htbp]
    \begin{center}
        \caption{Successful arrival-rate comparison results for a $5\times5\times4$ system configuration under Transpose traffic.} 
        \label{tb:arrival-tran}
        \begin{tabular}{|l|c|c|c|c|c|}
            \hline 
            Algorithm / Fault-rate&1\%&5\%&10\%&15\%&20\%\\ 
            \hline \hline 
            XYZ & 85\% & 46\% & 31\% & 14\% & 11\%\\ 
            \hline
            Hybrid-XYZ & 99\% & 68\% & 42\% & 25\% & 20\%\\ 
            \hline
            8-RW & 93\% & 82\% & 62\% & 44\% & 36\%\\ 
            \hline
            Odd-Even & 97\% & 84\% & 53\% & 42\% & 32\%\\ 
            \hline
            Hybrid-Odd-Even & 99\% & 92\% & 77\% & 62\% & 53\%\\ 
            \hline
            4N-FIRST & 96\% & 86\% & 68\% & 50\% & 37\%\\ 
            \hline
            4NP-FIRST & 100\% & 97\% & 89\% & 75\% & 63\%\\ 
            \hline
            LAFT-OASIS & 100\% & 100\% & 100\% & 99\% & 96\%\\
            \hline
            3D-FETO & 100\% & 100\% & 100\% & 99\% & 98\% \\
            \hline  
        \end{tabular} 
    \end{center}
\end{table}
\subsubsection{{Mean Time To Failure Improvement}}

\begin{figure*}[bht]
    \begin{center}
        \includegraphics[width=.9\linewidth]{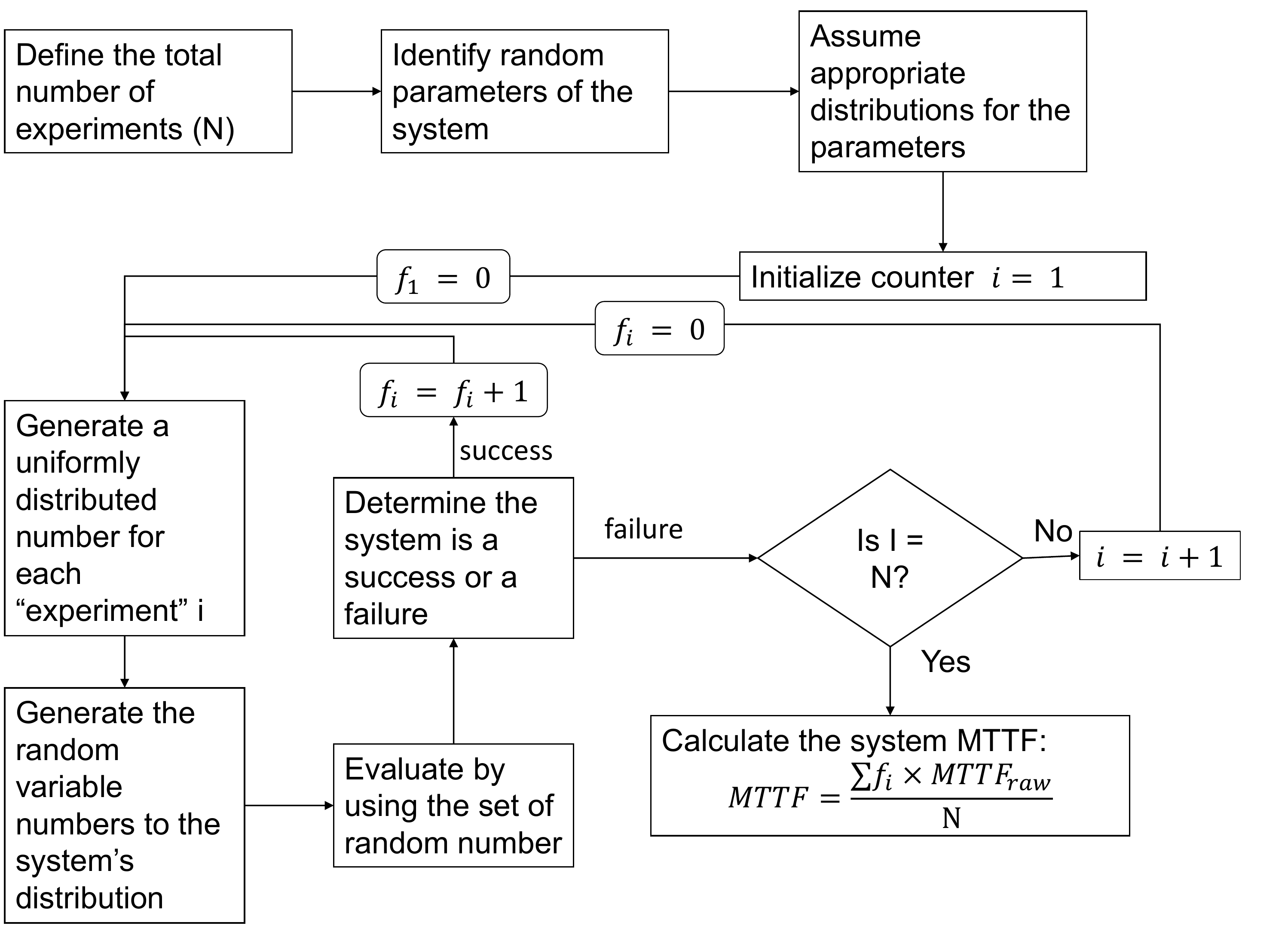}
        \caption{{MTTF simulation methodology.}}
        \label{fig:mc-sim}
    \end{center}
\end{figure*}

{Besides the arrival rate evaluation, we assessed our fault-tolerant system in terms of Mean Time To Failure (MTTF) improvement. We define a system at healthy if it operates correctly (100\% arrival rate, accurate fault detection and recovery function). Otherwise, the system is marked as \textit{failed}. To obtain more precise results, we use the net-list (gate-level) models from the complexity evaluation. Moreover, faults are not only injected to the fault-tolerance modules but they are also injected to other modules (controller, management module). Before the MTTF assessment, we first assume the original system has a natural fault rate: $\lambda_{raw}$. The MTTF value can be given as the following.}
\begin{equation}
MTTF_{raw} = \frac{1}{\lambda_{raw}}
\end{equation}
{To measure the MTTF value of the fault-tolerant system, we use a Monte-Carlo based simulation as shown in Figure}~\ref{fig:mc-sim}.{ At the beginning of the simulation, we define the number of experiments (N) and the fault models and distribution mechanisms. Faults will be generated in two types: soft errors (randomly occur within a clock period) and hard faults (occur from the beginning to the end of experiment). There are also two fault models: stuck-at ``0'' and stuck-at ``1''. Faults are injected to the dedicated gates which selected by a random generator. We use two distributions: (1) flat: randomly inject to any gate inside a router; (2) weight: more than 80\% of faults are injected to the fault-tolerant modules (buffer, crossbar, next-port-computing, switch-allocator). }
{ For each experiment $i$, we inject faults and examine the correctness of the system (data's accuracy, fault-tolerance configurations). Faults will be injected until the system is determined as failure. At the end of an experiment, the number of faults is recorded for the final process. 
To calculate the MTTF value of a system, the average number of faults is used in the following equation. }

\begin{equation}
    MTTF_{system} = \frac{\sum f_i \times MTTF_{raw}}{N}
\end{equation}
{In order to understand the efficiency of the fault-tolerance, the ratio of two MTTF values is used as in Equation}~\ref{eq:mttf-impv}.
\begin{equation} \label{eq:mttf-impv}
Improvement_{MTTF} = \frac{MTTF_{fault-tolerant}}{MTTF_{original}}
\end{equation}

{Because the raw fault rate depends on the technology parameters and the operating conditions, they will require a highly complex evaluation. To alleviate the complexity, we assume the fault-tolerant and original system have a similar raw fault rate. Therefore, the MTTF improvement can be obtained by Equation}~\ref{eq:aftf-impv} {where, \textit{AFTF} is average fault to failure.
}

\begin{equation} \label{eq:aftf-impv}
Improvement_{MTTF} = \frac{AFTF_{fault-tolerant}}{AFTF_{original}}
\end{equation}
\begin{table*}[htbp]
    \begin{center}
        \caption{{Average number of faults to failure.} }
        \label{tb:mttf-table}
        \begin{tabular}{|l|c|c|c|c|c|}
            \hline 
            Fault-Type & Distribution & Baseline router & SHER-3DR router & MTTF Improvement \\ \hline \hline 
            \multirow{2}{*}{Hard Fault} & Flat 	&	2.37 &	4.58 & 1.93\\ \cline{2-5}
            & Weighted	&   2.055 &	6.085 & 2.96 \\ \hline
            \multirow{2}{*}{Soft Error} & Flat 	& 17.928 & 26.770 & 1.49 \\ \cline{2-5}
            & Weight&4.037	&	21.492 & 5.32 \\ \hline
        \end{tabular} 
    \end{center}
\end{table*}
%

{Table}~\ref{tb:mttf-table}{ shows the average number of faults to failure after 1000 simulations. The test scheme is built to functionally verify the data communication and the fault-tolerance mechanisms. In the flat distribution, the proposed SHER-3DR enhances the MTTF of hard faults and soft errors by 1.93 and 1.49 times, respectively. With the weight distribution, the proposal shows more improvement since the faults focus on the fault-tolerant modules. SHER-3DR's hard fault tolerance is 2.96 times better the baseline OASIS router. In terms of soft error MTTF, SHER-3DR is 5.32 times better than the original router. In conclusion, we observe a significant improvement in terms of MTTF from our proposed mechanism. Along with the high arrival rates, we demonstrated the reliability enhancement of our system.}

%

\section{Conclusion and Future Work}  \label{sec:concl}
In this paper, we proposed a comprehensive fault tolerant 3D-Network-on-Chip (3D-NoC) system architecture for highly-reliable many-core Systems-on-Chips (SoCs), named 3D-FETO. The proposed system is based on two approaches. First, a comprehensive mechanism to handle both soft error and hard faults in a 3D-NoC router is proposed. The hard fault support is achieved by leveraging reconfigurable components to handle permanent faults in links, input buffers, and crossbars, while soft error tolerance is obtained via efficient and light-weight software redundancy that enables fault recovery in the router pipeline stages. In the second approach, the system can support a detection, diagnosis and recovery technique which makes it independent of any complex and costly testing mechanisms commonly found in conventional systems. 

Through extensive evaluation, we showed that the proposed 3D-FETO was able to recover efficiently from a significant number of soft and hard errors at different fault-rates, reaching up to 33\%. This means that 3D-FETO can provide up to a 98\% packet arrival rate even when almost one-third of its components have failed. Despite the performance degradation and hardware complexity penalty, we still consider that this overhead is acceptable. This is because we made sure that the system is still functional at high fault rates where previously proposed systems fail to deliver packets. As reliability constitutes one of the main challenges in future SoC design, we demonstrated that the proposed 3D-FETO can be used as a reliable and independent system capable of ensuring fault resiliency in worst case scenarios and that it can be adopted for mission critical applications where correct data delivery is primordial.  

As a future work, we are planning to investigate the faults within Through-Silicon-Vias of 3D-ICs/3D-NoCs to provide a sufficient fault-tolerance method for 3D-NoC systems. Moreover, the degradation factors of the reliability, such as thermal stress, operating voltages, design characteristics should be also studied.

\begin{acknowledgements}
This work is partially supported by Competitive Research Funding (CRF), The University of Aizu, Reference P-11 (2016), and JSPS KAKENHI Grant Number JP30453020.
This work is also supported by VLSI Design and Education Center (VDEC), the University of Tokyo, Japan, in Collaboration with Synopsys, Inc. and Cadence Design Systems, Inc. 
The first and the last authors in the author-list are the main contributors of this work. 
\end{acknowledgements}

\end{document}